\documentclass[journal=acs,manuscript=article]{achemso}
\usepackage[utf8]{inputenc}

\usepackage{mathptmx,graphicx}

\usepackage{soul}
\usepackage{wrapfig}
\usepackage{amsmath}
\usepackage{amssymb}
\usepackage[export]{adjustbox}
\usepackage{float}
\usepackage{chemformula}

\usepackage[colorlinks=true, allcolors=blue]{hyperref}
\usepackage[T1]{fontenc}
\usepackage{textgreek}
\usepackage{csquotes}
\usepackage{comment} 

\usepackage{wasysym} 
\usepackage{mathtools} 

\title{Phase Matching Free Sensing with Undetected Light Using a Nonlinear Thin-Film Metasurface}

\author{Toby Severs Millard}
\affiliation{Blackett Laboratory, Department of Physics, Imperial College London, London SW7 2AZ, UK}
\alsoaffiliation{Quantum Materials Group, Quantum Technologies Department, National Physical Laboratory, Teddington TW11 0LW, UK}
\email{t.severs-millard21@imperial.ac.uk}

\author{Nathan Gemmell}
\affiliation{Blackett Laboratory, Department of Physics, Imperial College London, London SW7 2AZ, UK}
\email{n.gemmell20@imperial.ac.uk}

\author{Ross C. Schofield}
\affiliation{Blackett Laboratory, Department of Physics, Imperial College London, London SW7 2AZ, UK}

\author{Mohsen Rahmani}
\affiliation{Advanced Optics \& Photonics Laboratory, Department of Engineering, School of Science \&
Technology, Nottingham Trent University, Nottingham NG11 8NS, United Kingdom}

\author{Alex S. Clark}
\affiliation{Blackett Laboratory, Department of Physics, Imperial College London, London SW7 2AZ, UK}
\alsoaffiliation{Quantum Engineering Technology Laboratories, University of Bristol, Bristol BS8 1QU, UK}

\author{Chris C. Phillips}
\affiliation{Blackett Laboratory, Department of Physics, Imperial College London, London SW7 2AZ, UK}

\author{Rupert F. Oulton}
\affiliation{Blackett Laboratory, Department of Physics, Imperial College London, London SW7 2AZ, UK}
\email{r.oulton@imperial.ac.uk}

\begin{document}

\begin{abstract}
In this article, we report classical sensing with undetected light using octave spanning stimulated four-wave mixing from a plasmonic metasurface. The bidirectional nonlinear scattering due to inherent reflections from such thin nonlinear materials modifies their operation within a nonlinear interferometer. The theoretical model for visibility accounting for such bidirectionality as well as pulsed illumination accurately predicts visibility in the system as a function of transmission in the near-infrared seed (idler) arm. Spectrally resolving the visible signal emission evaluates the total dispersion within the interferometer, highlighting the prospect of ultrafast sensing with undetected photons.

\end{abstract}

\maketitle

\subsection*{Introduction}
Nonlinear interferometry allows phase and amplitude information to be passed between light that only interacts with an object and light that is only detected.\cite{Lemos2014_QUIP_OG,Lahiri2015_QUIP_Theory,Cardoso_2018_ClassicalIUL,Shapiro2015_ClassicalIUL_Theory} 
The interferometric technique uses correlated photon beams created in a nonlinear material, either via difference frequency generation\cite{Cardoso_2018_ClassicalIUL} or parametric fluorescence\cite{Lemos2014_QUIP_OG}. 
When implemented with non-degenerate beams, there are wide-reaching applications where passing information between wavelengths is desirable; for example, where the technical and fundamental limitations of mid-infrared (MIR) detectors can be circumvented with silicon-based sensors\cite{Rogalski_2016_IRdetector_challenges,Lindner2020_FTIR_UP_low-gain,Lindner2023_MIRsensing_UP_SiDetect} for label-free biological imaging,\cite{Greaves2023_SNOM_intracell_labelFree,Kviatkovsky2020_QIUP_MIR} environmental monitoring,\cite{Myers2009_MIRsensing_OrganPollutant} and medical imaging.\cite{Amrania2018_Digistain,Contreras-Rozo2023_MIRspecDiabetes} 

Recent research has predominantly focused on the quantum mechanical variant of nonlinear interferometry; where spontaneous parametric down conversion (SPDC), typically created in a bulk nonlinear crystal (NLC), interferes by induced coherence without induced emission.\cite{Wang1991_InducedCoherenceWithout,Lemos2014_QUIP_OG,Kviatkovsky2020_QIUP_MIR,Basset2021_videoRateQIUP,Kviatkovsky2022_QIUP_MIR_Microscopy,Pearce2023_compact} 
Here, indistinguishable photon pairs are generated in two paths of an interferometer such that they occupy two overlapping beam paths, labelled here as the high energy signal and the low energy idler. 
Induced coherence without induced emission occurs due to the lack of "which path" information; an object placed in one of the idler paths, for example, introduces path knowledge that destroys interference. 
What differentiates quantum nonlinear interferometry from other quantum sensing techniques, such as ghost imaging\cite{ErkmanShapiro_2010_GhostReview},
is that the sensing information can be found by only detecting photons that never interacted with the object. 

Quantum nonlinear interferometry research has resulted in low cost and portable\cite{Pearce2023_compact} near-infrared (NIR) imaging systems at video rate\cite{Basset2021_videoRateQIUP,Pearce2024_VideoRate_QIUP} and MIR operation.\cite{Kviatkovsky2020_QIUP_MIR,Kviatkovsky2022_QIUP_MIR_Microscopy} 
While the operating frequency range is set by the properties of the more common NLCs, currently achieved at up to 4 \textmu m with 800 nm detection,\cite{Kviatkovsky2022_QIUP_MIR_Microscopy} more exotic NLCs are being explored for operation up to 10 \textmu m.\cite{Kumar2021_AgGaS2_photonPair,Paterova2022_Qspec_AgGaS2_8um,Mukaii2022_QFTIR_AgGaS2} 
However, crystals of order millimetres in length are required to generate sufficient levels of nonlinear emission through phase matching techniques. 
Alternative media that promise range extension are nonlinear metasurfaces; subwavelength thick, nanostructured arrays that enhance nonlinear processes by supporting electromagnetic resonances with high field localisation.\cite{Satiago-Cruz2021_LN_metasurface,Santiago-Cruz2022_GaAs_BIC_metasurface,Son2023_SPDC_BICbidirectional_MS,Zhang2024_photonIndistinguish_SiMetasurface} 
Metasurfaces can be tailor-made for operation at targeted frequencies and their subwavelength thickness relaxes any phase matching and absorption constraints of the resonant nonlinearity.\cite{Schulz2024_MetasurfaceRoadMap,Oulton2022_NPnews&views_thinNL} 
While metasurfaces outperform unpatterned films of the same thickness and material in regards to spontaneous photon pair emission rates,\cite{Satiago-Cruz2021_LN_metasurface,Son2023_SPDC_BICbidirectional_MS} they are currently insufficient for implementing quantum nonlinear interferometry. 
However, by stimulating emission -- and therefore operating in the classical regime -- metasurfaces become suitable nonlinear sources.

Classical nonlinear interferometry achieves the same effect as the quantum implementation, passing amplitude and phase information across wavelengths, but via stimulating the parametric fluorescence; for example, stimulating the idler while detecting the signal.\cite{Shapiro2015_ClassicalIUL_Theory,Cardoso_2018_ClassicalIUL,Huang2023_fibreStimFWMimaging} 
The advantages of the classical variant over the quantum mechanical are an improved signal-to-noise ratio and greater coherence of the bi-photon state.\cite{Cardoso_2018_ClassicalIUL,Kolobov_2017_ControllingInducedConherence}
Upconversion sensing also relies on stimulated nonlinear emission;\cite{Johnson2012_MIRfs_upcon,Ashik2019_MIRfs_upcon} however, classical nonlinear interferometry distinguishes itself with phase sensitivity, stronger robustness to noise, enhanced sensitivity at low amplitude, and immunity to incoherent backgrounds.

Here, we demonstrate classical nonlinear interferometry to achieve phase matching free sensing with undetected photons using a resonant gold plasmonic nonlinear metasurface, where four-wave mixing (FWM) is stimulated using femtosecond pulsed pump and seed illumination. 
In our case, the seed beam stimulates the interferometer's NIR idler mode to produce FWM in the visible signal mode. 
Amplitude and phase information of one path of the idler is observed in the visibility and phase of the signal mode interference. 
Operated as a folded Michelson interferometer, we study the influence of pump, seed and idler reflections combined with the bidirectional FWM emission of the metasurface. 
Unlike conventional nonlinear crystals, metasurface reflections cannot be eliminated due to their resonant nature. 
Despite these reflections and the high-order nonlinearity involved, we find that the homodyne advantage leads to a signal interference visibility of over 50\%. 
We also find signal interference independent of pump-seed delay outside the interferometer and discuss how this can be used for ultrafast sensing with undetected light. 
Although experimentally implemented with a metasurface, the theoretical framework is appropriate for any thin-film nonlinear source.

\subsection*{Results \& discussion} 
\subsubsection{Nonlinear interferometry with a thin-film source}
\begin{figure*}[ht!]
    \centering
    \includegraphics[width=0.885\linewidth]{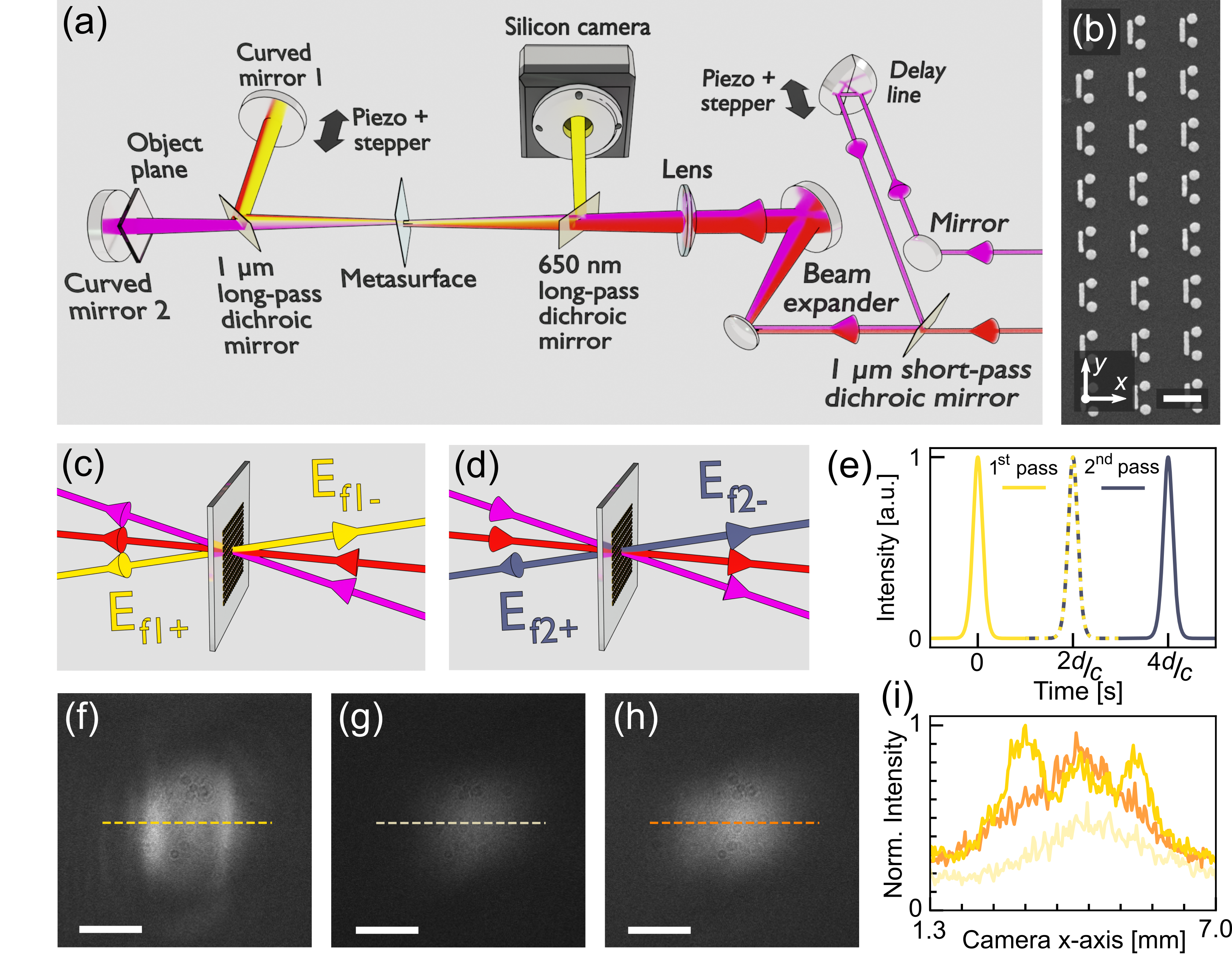}
    \caption{\textbf{The interferometer system, metasurface and interference} (a) The Michelson interferometer setup operating in the far field. The 835 nm pump (red) and 1500 nm seed (idler, magenta) pulsed lasers stimulate 580 nm FWM (signal, yellow). (b) SEM image the nonlinear metasurface comprised of two gold discs and a gold bar antenna per unit cell. The scale bar is 500 nm. (c, d) Graphical illustration of the bidirectional FWM (signal) emission at the first and second pass, respectively. The optical axes have been angularly displaced for clarity and do not indicate amplitude. (e) Temporal positions of FWM (signal) pulses created at each pass. The dashed lines represent $E_{f1+}$ and $E_{f2-}$. Distance $d$ is from the metasurface to curved mirror 1. (f, g, h) Intensity of FWM (signal) at the camera with no beam blocks in the interferometer arms, when beams are blocked at curved mirror 1, and when the seed (idler) is blocked at curved mirror 2, respectively. Scale bars are 2 mm. (i) Intensity at the camera along the dashed lines on (f, g, h). }
   
    \label{fig:1}
\end{figure*}

In the Michelson-style interferometer, Fig.~\ref{fig:1}(a), FWM (signal) is generated on two pump and seed passes through the nonlinear metasurface, shown in Fig.~\ref{fig:1}(b). Pulses at 835 nm (pump, red) and 1500 nm (seed, magenta) are overlapped in transverse space at a 1 \textmu m short-pass dichroic mirror, and in time using a piezo-controlled delay line in the seed path. Both beams are expanded to $\sim$10 mm in diameter and focused using a $20$ cm focal length lens to illuminate a $75\times 75$ \textmu m$^2$ metasurface across an areal overlap of 23.4 \textmu m diameter. 
The peak powers delivered by the pump and seed fields are 14.5 kW and 80.0 W, respectively. 
The metasurface is composed of gold antennas that are doubly resonant at the pump and seed wavelengths leading to enhanced electric field overlap at the incident wavelengths, generating FWM emission at 580 nm (signal, yellow) and 1500 nm (idler, magenta). 
The subwavelength thickness of the nonlinear metasurface relaxes out-of-plane phase matching, resulting in FWM (signal) in both directions on each pass, illustrated in Fig.~\ref{fig:1}(c, d). Combined with reflections of the various wavelengths from the metasurface, the nonlinear interference is therefore distinct from conventional nonlinear interferometry. 
From a single pass through the metasurface, the peak power of FWM (signal) emission is $4.81$ nW across a $1.17$ mm diameter spot at the camera. 
The photon conversion efficiency relative to the seed beam is estimated to be $1.18\times10^{-11}$. 
When the seed beam is blocked, no FWM (signal) is observed at the camera using typical settings of $800$ ms exposure time and $600 \times$ EM gain. 
Both the linear and nonlinear responses of the structures have been studied in greater detail in previous literature.\cite{Gennaro2016_BarandDiscsSHG,Gennaro2018_barDiscs_PulseChar,Yang2025_StimTomography_B&D}

We define the frequency domain pump and seed fields on the first pass of the metasurface as ${{\bf E}_{p1}=E_p(\omega)e^{i\phi_p(\omega)}\hat{x}}$ and ${{\bf E}_{s1}=E_s(\omega)e^{i\phi_s(\omega)}\hat{y}}$, respectively, where polarisation is relative to the axes marked on Fig.~\ref{fig:1}(b). FWM (signal) is generated as $\omega_f=2\omega_p-\omega_s$. After the metasurface, the seed (idler) is separated from the pump and FWM (signal) at a 1 \textmu m long-pass dichroic mirror, and a pair of identical curved mirrors with $20$ cm radius of curvature reflect all wavelengths back along their respective paths to the metasurface. On the second pass, the returning pump field at the metasurface is ${\bf E}_{p2}=\sqrt{T_p} {\bf E}_{p1} e^{2ik_p d}$, where $T_p$ is the metasurface pump transmission and $d$ is the optical path length of the pump and FWM (signal) arm of the interferometer. Meanwhile, the returning seed field is ${\bf E}_{s2}=\sqrt{T_s T}e^{i\gamma} {\bf E}_{s1} e^{2ik_s d^\prime}$, where $T_s$ is the seed transmission of the metasurface, $T$ is the transmission coefficient and $\gamma$ is the phase shift of the object to be sensed, and $d^\prime$ is the optical path length of the interferometer’s seed (idler) arm. FWM (signal) emission at each pass occurs in both $+$ and $-$ directions. 
The FWM field amplitude on the first pass is $\sigma_\pm(\omega)=\alpha_\pm E_{p}^2 E_{s}^*$, where $\alpha_\pm$ indicates the strength of the nonlinear process, which is related to the nonlinear $\chi^{(3)}_{\mu xxy}$ tensor elements for the process. 
The spontaneous emission is assumed to be negligible in comparison to the stimulated emission. 
FWM (signal) on the second pass is reduced by the transmission and phase at each wavelength.
We thus identify four FWM (signal) amplitudes at the metasurface per incident pump and seed pulse, illustrated and labelled in Fig.~\ref{fig:1}(c, d), and evaluate their resulting amplitude and phase at the camera to be
\begin{equation}
E_{f1-} = \sigma_- = \alpha_{-}E_{p1}^{2}E_{s1}^{*}e^{ik_{f}l} 
\end{equation}
\begin{equation}
E_{f1+} = \sqrt{T_f} \sigma_+ e^{2ik_f d}
\end{equation}
\begin{equation}
E_{f2-} =\sqrt{TT_p^2 T_s}e^{i\gamma}  \sigma_- e^{2ik_f d+i\psi} 
\end{equation}
\begin{equation}
E_{f2+} =\sqrt{TT_fT_p^2 T_s} e^{i\gamma} \sigma_+ e^{4ik_f d+i\psi} 
\end{equation}

\noindent where $T_f$ is the metasurface FWM (signal) transmission, $2k_f (d^\prime-d)=\psi$, is the phase difference between the interferometer's arms. 
The normally incident and expanded pump and seed beams ensure that the nonlinear emission is nearly in-phase across the metasurface, which is equivalent to matching the in-plane momenta of the various fields, such that the longitudinal momentum is $2k_p-k_s=k_f$. 
The FWM (signal) is isolated from the pump and seed and directed towards a silicon camera using a $650$ nm long-pass dichroic mirror. 
The camera thus integrates over three FWM pulses per input pump and seed pulse with relative time delays of $0$, $2d/c$ and $4d/c$, as illustrated in Fig.~\ref{fig:1}(e).
Since the delay between distinct FWM pulses, $2d/c$, is greater than the FWM (signal) pulse duration, no interference occurs between them, and the camera measures a time-averaged intensity

\begin{equation} \label{TimeAvgIntensity}
D=|E_{f1-} |^2+|E_{f1+}+E_{f2-} |^2+|E_{f2+}+\sqrt{R_f} E_{f1+} e^{2ik_f d} |^2.
\end{equation}

The third term in Eq.~(\ref{TimeAvgIntensity}) is an interference of the forward scattered FWM (signal) from the second pass, $E_{f2+}$, with the forward scattered FWM (signal) from the first pass reflected by the metasurface, $\sqrt{R_f} E_{f1+}$, where $R_f$ is the metasurface FWM reflectivity.
Linear reflections of the pump and seed beams by the metasurface have been omitted due to nonlinear scaling of FWM emission.
Curved mirror 1 is connected to a second piezo-controlled stage, allowing the phase difference, $\psi$, to be varied. 
The detected signal normalised to the forward emitted FWM (signal) emission signal, $\sigma_+^2$, is 

\begin{equation}
N=D/\sigma_+^2=R_f T_f+(\eta^2+T_f)(1+\kappa T)+2(\eta+\sqrt{T_f R_f})\sqrt{\kappa TT_f}\cos(\gamma+\psi),
\end{equation}

\noindent where $\kappa=T_p^2 T_s$ is the ratio of total FWM (signal) from the two passes and $\eta=\sigma_-/\sigma_+$ is the ratio of bidirectional emission. 
A single frame of the FWM interference at the camera can be seen in Fig.~\ref{fig:1}(f). 
When a beam block is placed in front of curved mirror 1, the forward emitted first pass, $E_{f1+}$, is no longer detected and the pump beam does not generate FWM (signal) at the second pass, $E_{f2\pm}=0$, so no interference is observed in Fig.~\ref{fig:1}(g), and the camera measures $\sigma_-^2$. 
When the beam block is placed in front of curved mirror 2, the seed no longer stimulates second pass emission, $E_{f2\pm}=0$, and interference is destroyed in Fig.~\ref{fig:1}(h), so the camera records $\sigma_-^2+(1+R_f)T_f\sigma_+^2$. 
Line plots in Fig.~\ref{fig:1}(i) clarify interference and its subsequent breakdown in Fig.~\ref{fig:1}(f-h). 
A video of the interference fringes as $\psi$ is varied can be found in the Supporting Information (SI). 
By characterising the linear transmissivity and reflectivity of each field, shown in Table I of the SI, we can evaluate the theoretical performance of the metasurface nonlinear interferometer. The expected visibility is $V=(N_{max}-N_{min})/(N_{max}+N_{min})$ such that

\begin{equation} \label{Eq_visibility}
 V = \frac{2\sqrt{\kappa T T_f}(\eta+\sqrt{T_fR_f})}{R_fT_f+(\eta^2+T_f)(1+\kappa T)}.
\end{equation}

At \(T=100\%\), the theoretical visibility predicts a maximum of 55.3 $\pm$ 0.5\% when applying $\eta$ extracted from the images in Fig.~\ref{fig:1}(g) and (h). For reference, a conventional nonlinear crystal with ideal anti-reflection coatings placed within this interferometer would only have two interfering FWM (signal) fields, $D_0=\sigma_0^2 |E_{f1+}+E_{f2-} |^2$; which for collinear phase-matched emission, implies $\sigma_\pm=\sigma_0$, such that $N_0=1+T+2\sqrt{T}\cos(\gamma+\psi)$, with a maximum visibility, $V_0=2\sqrt{T}/(1+T)\mapsto 1$ in the case of no object. A more detailed derivation of the visibility can be found in Section SII of the SI. 
We note that the reduction in the maximum visibility when using a metasurface source is predominantly due to the $E_{f1-}$ field. 
The FWM (signal) fields can be directionally filtered in a Mach-Zehnder interferometer to achieve a maximum visibility that is theoretically equivalent to conventional nonlinear interferometers.
Alternatively, tilting the metasurface to misalign nonlinear reflections also maximises visibility while simultaneously allowing the nonlinear emission to be monitored.
These configurations are discussed theoretically in Sections SIII and SIV of the SI.

\subsubsection{Sensing through opaque media}
\begin{figure*}[ht!]
    \centering
    \includegraphics[width=0.95\linewidth]{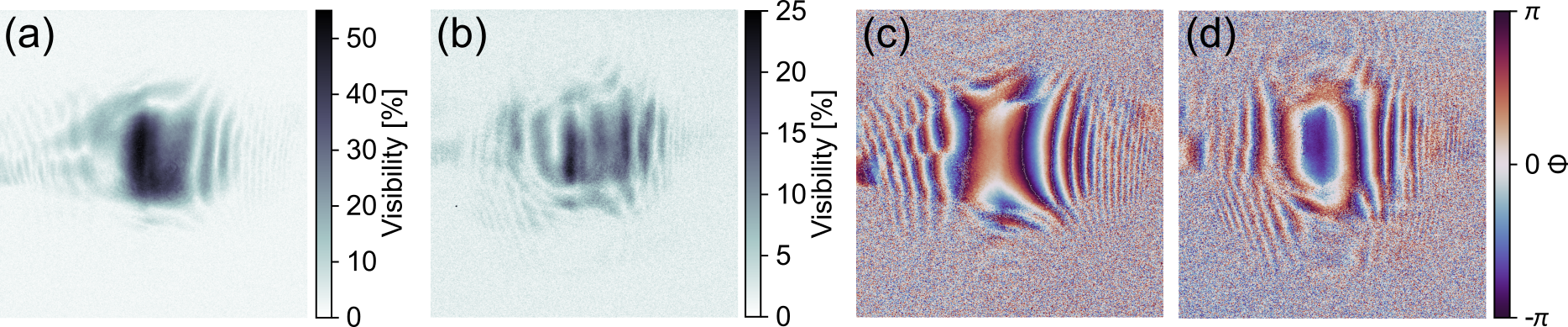}
    \caption{\textbf{Visibility and phase of the FWM (signal) interference, centred at 580 nm, with and without a 240 \textmu m silicon window in the seed (idler) arm of the interferometer}. Experimental visibility at the camera with no additional material in the seed (idler) path (a) and with the silicon window (b). The corresponding phase of interference, without (c), and with (d), the silicon window in the seed (idler) arm. 
   } \label{fig:2}
\end{figure*}

We compare the theoretical prediction to the experimental visibility in Fig.~\ref{fig:2}(a) with a maximum of 55.0\% $\pm$ 1.7\%. Here, the visibility is calculated by fitting $N=D(1+V\mathrm{cos}(\phi))$ to the interference at each pixel on the camera, where $\phi$ is the phase and $D$ is the DC intensity. 
As an interferometric technique, the corresponding phase map, Fig.~\ref{fig:2}(c), is simultaneously extracted. The saddle-like phase pattern is characteristic of interference between two fields; one diverging and the other converging. Ultimately, this condition resulting in maximised visibility is due to the same refractive lens focussing both the pump and seed to different focal points. 

An advantage of non-degenerate imaging and sensing with undetected light is the ability to observe at one wavelength while exploiting the optical properties of light-matter interaction at another. To highlight this effect, we apply the same process of extracting visibility and phase information to interference data collected through an anti-reflection coated 240 \textmu m piece of polished silicon placed at the object plane. At 580 nm, the silicon window is opaque, but at 1500 nm it is measured to have a single pass transmission of 91.3 $\pm$ 0.9\%, which leads to a theoretical maximum visibility of 52.3 $\pm$ 0.5\% in Fig.~S5(b) of the SI. 
In the experiment, Fig.~\ref{fig:2}(b), the visibility has dropped to a maximum of 24.4 $\pm$ 1.7\%. 
The path length introduced by the silicon window is $\sim1.2$ mm; greater than the coherence length of interference. 
Re-establishing temporal overlap is likely to have reduced nonlinear emission on the second pass, and in turn, reduced visibility. 
A further consequence of this realignment is that a direct measure of the phase change cannot be taken from Fig.~\ref{fig:2}(c) and (d). However, the clear pattern in (d) shows phase information is retrieved despite the reduced visibility. Additionally, the change in phase pattern suggests the focal points of the two passes have reversed from the alignment in (c).

\subsubsection{The metasurface's optical properties \& single-mode operation.}
\begin{figure*}[ht!]
    \centering
    \includegraphics[width=0.9\linewidth]{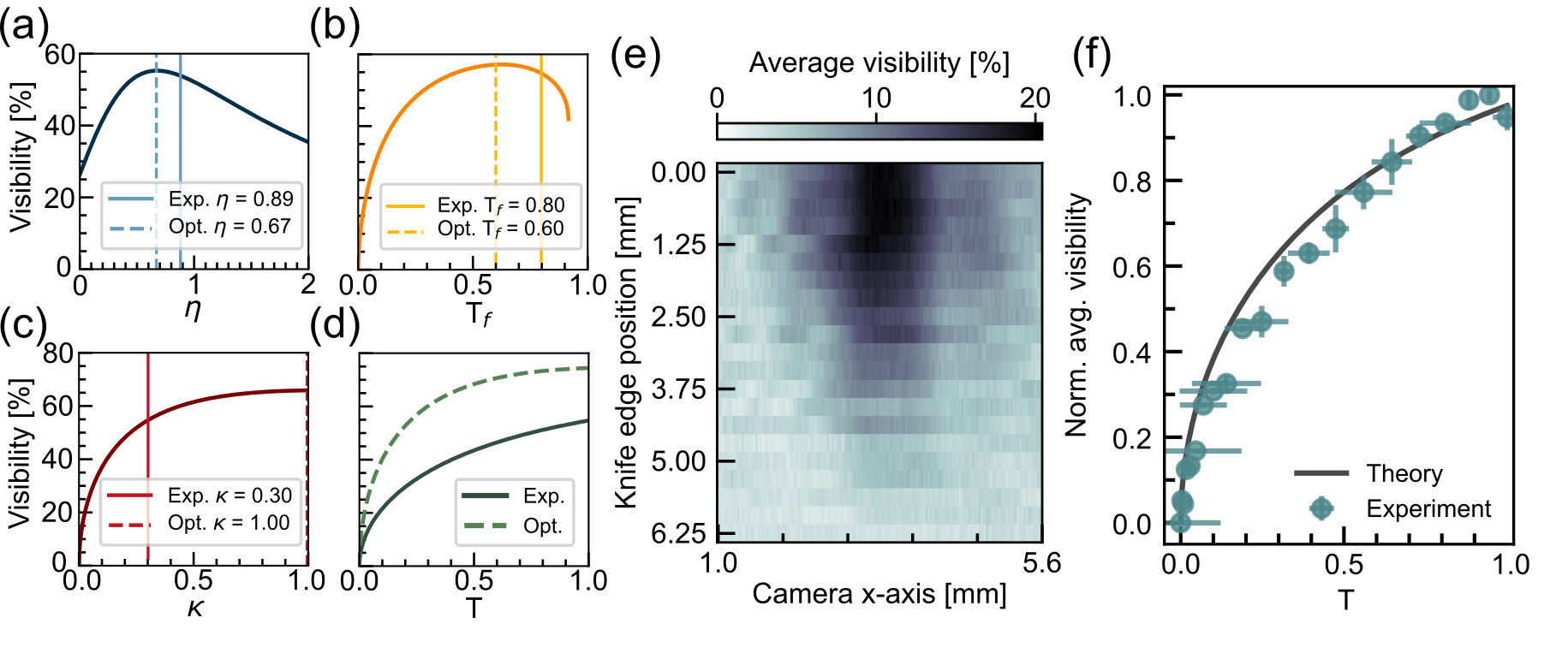}
    \caption{{ \bf Visibility as functions of the optical properties of the metasurface and transmission in the seed (idler) arm}. Theoretical dependence of visibility on (a) $\eta$, (b) T$_f$ and (c) $\kappa$. The experimentally measured and theoretically optimised metasurface parameters are marked by the solid and dashed vertical lines, respectively. Constant parameters are held at the values in Table SI of the SI. (d) Dependence of visibility on seed (idler) transmission, $T$, for the experimental measured (solid curve) and theoretically optimised (dashed curve) metasurface parameters. (e) Experimental average visibility measured at knife edge positions that reduce seed (idler) transmission. The knife edge is brought along the horizontal axis of the object plane and visibility is averaged across three rows of pixels. (f) Experimental visibility as the knife edge is introduced, averaged across all pixels and normalised to the unobstructed visibility. The theoretical curve uses the experimentally measured optical properties.
     }
    \label{fig:3}
\end{figure*}

The derived visibility is dependent on the optical properties of the nonlinear metasurface. Visibility as a function of $\eta$, the ratio of nonlinear emission ($\sigma_-$/$\sigma_+$), shown in Fig.~\ref{fig:3}(a), peaks at $\eta$ = 0.67, offset from 1 due to a trade-off between the two sets of interfering fields in Eq.~(\ref{TimeAvgIntensity}).
The first interfering term depends only on T$_f$ while the second depends on both T$_f$ and R$_f$. 
Therefore, maximising T$_f$ brings the upper limit of visibility down in Fig.~\ref{fig:3}(b), while bringing the optimal $\eta \mapsto 1$. Consequently, further optimisation could be achieved by iteratively tuning $\eta$ and T$_f$ (R$_f$).
The ratio of total FWM (signal) emission on the second pass compared to the first is represented by the balancing term $\kappa$. As both interfering terms in Eq.~(\ref{TimeAvgIntensity}) depend on emission from each pass, visibility is maximised at $\kappa=1$ in Fig.~\ref{fig:3}(c). 
Fixing the optical parameters at the experimental and optimised values in Fig.~\ref{fig:3}(a-c) leads to the visibility curves in Fig.~\ref{fig:3}(d) as functions of the seed (idler) transmission. 
With this being said, an optimised metasurface may be difficult to engineer, requiring perfect transmission at the pump and seed wavelengths while balancing the bidirectionality of nonlinear emission with the reflection and transmission of FWM (signal).

In the stimulated FWM process at the metasurface, the pump and seed modes generate a single idler mode. 
The resulting interferometer is therefore a single-mode sensor and does not offer imaging capability. 
Any additional modes created at an object do not interfere with the single mode of the first pass. 
By scanning a knife edge horizontally across the object plane of the interferometer we can measure a reduction in visibility across the horizontal (x-axis) of the camera, as shown in Fig.~\ref{fig:3}(e). 
The knife edge starts by not obstructing the seed (idler) beam at 0.00 mm to fully obstructing it at 6.25 mm, and visibility is calculated at each knife edge position.
The profile of the FWM mode remains the same shape, with horizontal visibilities falling together as the seed (idler) is partially to fully obstructed. 
Fig.~\ref{fig:3}(f) displays the normalised visibility averaged across the entire camera sensor area for the same experiment. 
The knife edge position has been converted to normalised seed (idler) transmission, $T$, to identify the single-mode object transmission. 
The theoretical model predicts visibility using the experimentally measured optical parameters of the metasurface (Table I of the SI). 
The visibility is normalised to account for averaging across the camera sensor.
\subsubsection{Spectroscopy \& dispersion}
\begin{figure*}[ht!]
    \centering
    \includegraphics[width=0.999\linewidth]{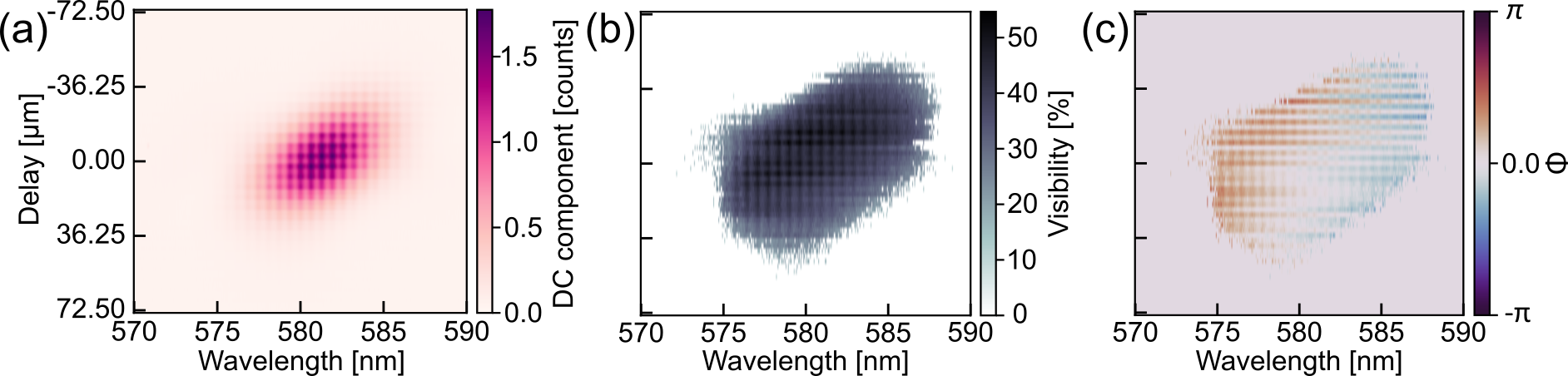}
    \caption{\textbf{2D FWM interference spectrograms for wavelength against pump-seed delay position.} (a) Intensity of the FWM (signal) DC component indicating the chirp of spectral components. (b) Spectrally resolved FWM (signal) visibility as a function of pump-seed delay,  measured by scanning the interferometer idler mirror at each delay position. (c) The spectrally resolved accumulative phase from all beams as a function of pump-seed delay showing the small intrinsic dispersion of the interferometer. For (b) and (c), data points extracted with a coefficient of determination, R$^2< 0.7$, have been removed.
     }
    \label{fig:4}
\end{figure*}

As the interferometer is operated with ultrafast pulses, spectral visibility and phase information are available within the FWM emission bandwidth. 
To access this information, the silicon camera is replaced with a spectrometer and appropriate optics. 
The piezo-controlled seed delay line, positioned before the 1 \textmu m short-pass dichroic, is scanned across a 145 \textmu m (483 fs) range with 2.5 \textmu m (8.3 fs) steps, centred at 0 \textmu m where the pump and seed are at their best temporal overlap that maximises the FWM (signal) emission in Fig.~\ref{fig:4}(a). 
The FWM (signal) spectrum blue shifts as the delay position moves from negative to positive, indicating chirp in the pump and seed pulses. The fine fringe pattern in the spectral data originates from the 175 \textmu m thick borosilicate substrate that the metasurface sits on, which acts as a Fabry-Perot cavity. At each pump-seed delay position, the piezo-controlled curved mirror inside the interferometer is scanned across three interference fringes to extract the visibility and phase in Fig.~\ref{fig:4}(b) and (c). 

Despite the clear peak in FWM (signal) emission in Fig.~\ref{fig:4}(a), the visibility in Fig.~\ref{fig:4}(b) shows a flatter profile, and a continuous phase relationship can be extracted across the entire spectral mode. 

The spectral shift in Fig.~\ref{fig:4}(c) shows the difference in phase between FWM (signal) generated at the first and second pass. Therefore, the phase shift comes from the combined dispersion experienced by the seed (idler), pump, and FWM (signal) after the first pass. All three pulses travel twice through the 175 \textmu m borosilicate substrate of the metasurface, while the seed (idler) also travels twice through a 1 mm thick dichroic mirror nominally positioned at 45 degrees to the incoming beams. 
Averaging the phase shift in Fig.~\ref{fig:4}(c) across all delay positions from 575 to 585 nm gives $-1.56 \pm 0.29$.
Importantly, the interferometer is only sensitive to variations in the object's phase, while changes in the FWM (signal) phase for different delay positions have no effect on the phase variation, as seen in Fig.~\ref{fig:4}(c). 
Interestingly, with the addition of a second pulsed beam to illuminate the object plane, a Fourier transform of the interferometer visibility and phase would reveal ultrafast transients of the object transmission with a resolution less than the pulse duration.\cite{Kaur2024PulsedQFTIR} 
With this in mind, the pulsed modes are expected to preserve their transform limited nature in our system; experiencing no temporal broadening from group velocity mismatch or phase matching bandwidth narrowing associated with bulk nonlinear material.


In conclusion, we have demonstrated the operation of a nonlinear interferometer that uses a nonlinear plasmonic metasurface as a source to generate signal and idler photons through stimulated FWM to implement classical sensing with undetected light. The ability to pass amplitude and phase information from the NIR (1500 nm) to the visible (580 nm) is demonstrated by sensing through a 240 \textmu m silicon window while detecting the emission and its phase on a silicon camera. Meanwhile, a knife edge was used to modulate the single-mode transmission of the interferometer. A theory of the interferometer based on any thin nonlinear source with intrinsic reflections accurately predicts the experimental visibility as a function of NIR seed (idler) transmission.


The single-mode nature of interference -- brought about by seeding the FWM (signal) with single mode lasers -- provides no spatial information for imaging. 
However, additional modes could be generated before the first pass through the metasurface. 
Alternatively, imaging could be achieved by adapting the interferometer to operate in real space, where the object plane is imaged onto the metasurface, which is in turn imaged onto the camera. 
Similarly, for spectroscopy, the system can utilise the broad spectral range of the resonant metasurface, either by tuning the seeding wavelength or illuminating with a broadband seed beam.

The spectral breadth of the signal and idler generated through pulsed illumination allows for spectrally resolved visibility and phase to be measured, assessing dispersion within the interferometer. Additionally, pulsed illumination opens up avenues for ultra-fast pump-probe spectroscopy with undetected photons, such as lifetime measurements for molecular excited states and vibrations.

In combination, pulsed illumination and stimulated nonlinear emission reduce the requirement for millimetre-scale exotic crystals and phase matching. 
The nonlinear interferometer system introduced here has the potential to reach any wavelength range where a laser and metasurface resonances are spectrally overlapped. Therefore, operation deep in the MIR ($\sim$10 \textmu m) is conceivable, where micron-scale resonators are easy to fabricate and pulsed lasers are readily available.
Additionally, the interferometer model developed is independent of the thin-film source used, including thin films of nonresonant nonlinear crystals, allowing the operating wavelengths to be defined solely by the pump and seed beams.

\subsection*{Methods} \label{sec:methods}

\subsubsection*{Sample fabrication}
The metasurface was fabricated by electron beam lithography on a 175 \textmu m thick borosilicate glass substrate. The substrate was coated with the positive resist polymethyl methacrylate (PMMA) and baked at 180$\mathrm{^{o}}$C for at least 180 s. Nanoantenna geometries are defined by electron beam exposure at 20 keV using a 10 \textmu m aperture. The PMMA was developed in a 3:1 isopropyl alcohol (IPA) to methyl isobutyl ketone (MIBK) solution, and rinsed for 30 s in IPA to halt the development. A 1.5 nm adhesion layer of Cr was deposited by thermal evaporation onto the substrate, followed by 40 nm of Au (Angstrom Engineering Amod). PMMA and excess Cr/Au were removed by leaving the substrate in acetone for at least 24 hours and oxygen plasma ashing.

\subsubsection*{Optical measurements}
Optical measurements were taken using a seed laser at 1500 nm ($\lesssim$ 228 fs) generated by a Coherent Chameleon Compact OPO, pumped by a Coherent Chameleon Ultra II at 835 nm ($\lesssim$ 146 fs). 
The seed was sent to a piezo-controlled delay line before being combined with the pump at a 1 \textmu m short-pass dichroic mirror (Thorlabs DMSP1000). 
Both beams were passed through a $\times4$ reflective beam expander (Thorlabs BE04R) and focused onto the sample using a 20 cm focal length refractive lens. 
After the sample, the transmitted pump and forward generated FWM (signal) were separated from the seed (signal) by a 1 \textmu m long-pass dichroic mirror (Thorlabs DMLP1000). 
Identical curved mirrors, each with a 20 cm radius of curvature, reflected the respective beams back to the sample. 
The pump and FWM (signal) path lengths were varied by a piezo-controlled stage connected to the curved mirror in their path.  
A 650 nm long-pass dichroic mirror (Thorlabs DMLP650) isolated the FWM (signal) and directed it onto an EM-CCD camera (Hamamatsu C9100-23B ImagEM X2). 
FWM (signal) emission and interference were recorded using $800$ ms exposure time and an EM gain of $600\times$.

For spectrally resolved measurements, the camera was replaced with a 20 cm focal length lens to collimate the FWM (signal). A pinhole was used to select the centre of the interference pattern that was subsequently focused onto the entrance slit of the spectrometer (Acton SP2300) and detected on a silicon-based CCD. 

\subsubsection*{Data availability}
Data is available upon request.

\subsubsection*{Supporting information}
The supporting information contains; experimentally measured optical parameters of the metasurface, a more detailed derivation of the theory included in this article, the theory behind two additional interferometer configurations, and calculations of predicted visibility for our system. A video of the interference pattern as the phase is varied is also included.

\subsubsection*{Acknowledgments}
T.S.M. acknowledges the support of the UK government Department for Science, Innovation and Technology through the UK National Quantum Technologies Programme and the National Physical Laboratory. This work is supported by the UK Quantum Technology Hubs in Sensing Imaging and Timing EP/Z533166/1 and Quantum Imaging EP/T00097X/1 and by the UK Engineering and Physical Sciences Research Council (EPSRC), through the Reactive Plasmonics EP/M013812/1 and Catalysis Plasmonics EP/W017075/1 Programme Grants. M.R. acknowledges support from the UK Research and Innovation Future Leaders Fellowship (MR/T040513/1). A.S.C. acknowledges support from The Royal Society (URF/R/221019, RF/ERE/210098, RF/ERE/221060).

\subsubsection*{Author Contributions}
M.R. fabricated the metasurface.
T.S.M. collected optical measurements. 
R.F.O., T.S.M. and N.G. developed the theory. 
T.S.M. wrote the paper with contributions from all co-authors. 
R.F.O., N.G., A.S.C. and C.C.P. conceived and supervised the project.

\bibliography{FWM_bibliography}

\end{document}


\maketitle

 \section{Experimental optical parameters}
Transmission and reflection measurements were taken with the same tunable pulsed laser used for the main experiment at 835 nm and 1500 nm for the pump and seed (idler), respectively.
For measurements at 580 nm, the seed was tuned to 1160 nm and passed through a frequency doubler. 
The FWM (signal) emission in each direction, $\sigma_\pm$, was extracted from the intensities at the camera shown in Fig.~\ref{fig:S2}(a) and (b) (Fig.~1(g) and (h) of the main text), using constants from Table \ref{table:S1}. 
The visibility in Table \ref{table:S1} was calculated from all extracted parameters using Eq.~(7) from the main text (Eq.~(\ref{SI_visibility} below) and represents an average across the detector. 
\begin{table}[h!]
\begin{center}
\begin{tabular}{ l c c c }
 Parameter & Label & Value & Uncertainty ($\pm$)\\ 
 \hline
 Metasurface pump transmission & $T_p$ & 0.784 & 0.002 \\  
 Metasurface seed transmission & $T_s$ & 0.47 & 0.01 \\ 
 Metasurface FWM transmission & $T_f$ & 0.80 & 0.03 \\ 
 Metasurface FWM reflection & $R_f$ & 0.118 & 0.002 \\ 
 Normalised FWM emission in + direction & $\sigma_+$ & 0.530 & 0.002 \\ 
 Normalised FWM emission in - direction & $\sigma_-$ & 0.469 & 0.001 \\
 FWM emission ratio ($\sigma_-/\sigma_+$)  & $\eta$ & 0.885 & 0.004 \\
 Average visibility & $V$ & 0.54 & 0.01 
\end{tabular}
\end{center}

\caption{ \justifying{\bf Experimentally measured optical parameters of the metasurface used to calculate visibility.} }
\label{table:S1}
\end{table}

\section{Analytical visibility derivation} \label{SI_SEC_VisDerivation}

\begin{figure}[ht!]
    \centering
    \includegraphics[width=0.90\linewidth]{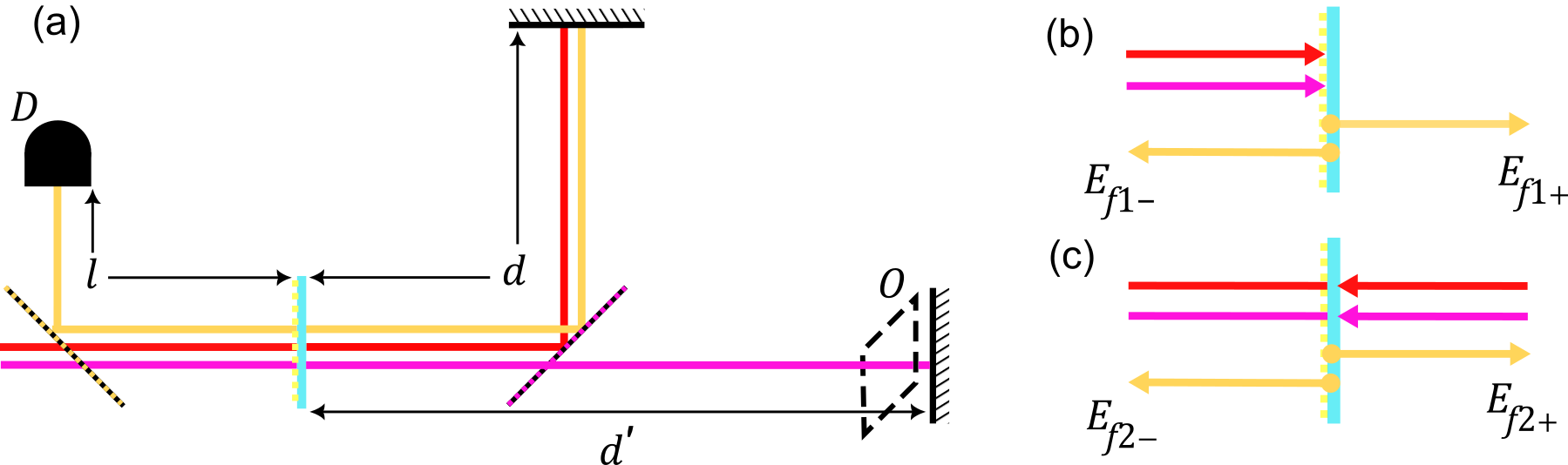}
    \caption{\justifying {\bf A simplified schematic of the interferometer system shown in Fig.~1(a) of the main text, to support the derivation below.} 
    (a) Beam paths of the pump (red), seed (idler, magenta) and FWM (signal, yellow) with the path lengths labelled.
    Due to the relaxation of out-of-plane phase matching requirements, FWM signal is emitted in both the $+$ and $-$ directions at the (b) first and (c) second pass. 
    This, combined with pulsed illumination, creates temporal windows where interference can occur.   }
    \label{fig:S1}
\end{figure}
At the first pass through the metasurface, the amplitude and phase of the pump and seed fields are described in the frequency domain, respectively, by 
\begin{equation} \label{SI_pump1}
   E_{p1} = E_{p}(\omega)e^{i\phi_p(\omega)} \textrm{ and}
\end{equation}
\begin{equation} \label{SI_seed1}
   E_{s1} = E_{s}(\omega)e^{i\phi_s(\omega)} \textrm{.}
\end{equation}
Here, $E_{p,s}(\omega)$ and $\phi_{p,s}(\omega)$ are the frequency dependent amplitude and phase of the pump and seed. 
After travelling the paths shown in Fig.~\ref{fig:S1}, the fields at the second pass become
\begin{equation} \label{SI_pump2}
   E_{p2} = t_{p}E_{p1}e^{2ik_{p}d} \textrm{ and}
\end{equation}
\begin{equation} \label{SI_seed2}
   E_{s2} = t_{s}te^{i\gamma}E_{s1}e^{2ik_{s}d^{\prime}} \textrm{.}
\end{equation}
Where $t_{p,s}$ are the pump and seed transmission amplitudes through the metasurface, and it is assumed the dichroic mirror is perfectly reflective for the pump and FWM, and is perfectly transmissive for the seed (idler).
The seed (idler) beam passes through an object or material at $O$ with transmission $t$, inducing a phase shift of $\gamma$.

Due to the subwavelength thickness, the phase matching requirements are relaxed and FWM (signal) is emitted out of the metasurface plane in both directions. 
However, due to the different refractive indices of the sub- and superstrate, the strength of emission is not equal in each direction relative to the metasurface (+ and -). 
We define FWM (signal) by the input beams,
\begin{equation} \label{SI_FWMstrength}
 E_{f1\pm,f2\pm} = \alpha_{\pm}E_{p1,p2}^{2}E_{s1,p2}^{*}  \textrm{,}
\end{equation}
where $\alpha_\pm$ indicates the strength of the nonlinear process, which is related to the nonlinear $\chi^{(3)}_{\mu xxy}$ tensor elements for the process. 
We are working under the assumption of undepleted beams, such that after FWM generation the pump and seed amplitudes are unchanged.

FWM fields generated at each pass and each direction are then evaluated at the detector using Eqs.~(\ref{SI_pump2}-\ref{SI_FWMstrength}) and path distances within the interferometer,
\begin{equation} \label{SI_f1-}
 E_{f1-} = \alpha_{-}E_{p1}^{2}E_{s1}^{*}e^{ik_{f}l}
\end{equation}
\begin{equation} \label{SI_f1+}
 E_{f1+} = \alpha_{+}t_{f}E_{p1}^{2}E_{s1}^{*}e^{ik_{f}(2d+l)}
\end{equation}
\begin{equation} \label{SI_f2-}
 E_{f2-} = \alpha_{-}E_{p2}^{2}E_{s2}^{*}e^{ik_{f}l} = \alpha_{-} te^{i\gamma}t_{p}^{2}t_{s}E_{p1}^{2}E_{s1}^{*}e^{ik_{f}l-2ik_{s}d^{\prime}+4ik_{p}d}
\end{equation}
\begin{equation} \label{SI_f2+}
 E_{f2+} = \alpha_{+}E_{p2}^{2}E_{s2}^{*}e^{ik_{f}(2d+l)} = \alpha_{+}t_{f}te^{i\gamma}t_{p}^{2}t_{s}E_{p1}^{2}E_{s1}^{*}e^{ik_{f}(2d+l)-2ik_{s}d^{\prime}+4ik_{p}d} \textrm{.}
\end{equation}
Here, $t_f$ is the transmission amplitude of the metasurface at the FWM wavelength. 
To simplify the above equations, we introduce the average nonlinear amplitude in each direction at either pass, $\sigma_\pm$,
\begin{equation} \label{SI_NIprocess_phase}
\alpha_{\pm}E_{p}^{2}E_{s}^{*}e^{ik_{f}l}e^{i(2\phi_p(\omega)-\phi_s(\omega))} = \widetilde{\sigma}_{\pm}(\omega)e^{i\phi_f(\omega)} = \sigma_{\pm} \textrm{,}
\end{equation}
where $\widetilde{\sigma}_{\pm}(\omega)$ represents the frequency-dependent nonlinear amplitude. We then consider the conservation of energy and momentum.
\begin{equation} \label{SI_E&K_conserve}
 2\omega_p - \omega_s = \omega_f \Rightarrow 2k_p - k_s = k_f \textrm{,}
\end{equation}
and relate the path lengths of the interferometer's arms
\begin{equation} \label{SI_pathDiff}
 d^{\prime} = d - \delta 
\end{equation}
\begin{equation} \label{SI_phaseShift}
 2k_s\delta = \psi \textrm{.}
\end{equation}
$\delta$ is the path difference between the interferometer arms that leads to a phase difference of $\psi$ on the seed (idler).
Substituting Eqs.~(\ref{SI_NIprocess_phase}-\ref{SI_phaseShift}) into Eqs.~(\ref{SI_f1-}-\ref{SI_f2+}) leads to
\begin{equation} \label{SI_f1-_simple}
 E_{f1-} = \sigma_{-}
\end{equation}
\begin{equation} \label{SI_f1+_simple}
 E_{f1+} = t_{f}\sigma_{+}e^{2ik_{f}d}
\end{equation}
\begin{equation} \label{SI_f2-_simple}
 E_{f2-} = tt_{s}t_{p}^{2}\sigma_{-}e^{i\gamma}e^{2ik_{f}d+i\psi}.
\end{equation}
\begin{equation} \label{SI_f2+_simple}
 E_{f2+} = tt_{f}t_{s}t_{p}^{2}\sigma_{+}e^{i\gamma}e^{4ik_{f}d+i\psi}
\end{equation}
Across Eqs.~(\ref{SI_f1-_simple}-\ref{SI_f2+_simple}) there are three time/phase windows at which FWM (signal) reaches the detector, separated by the time taken for light to traverse the interferometer $0$, $2k_{f}d$ and $4k_{f}d$. 
The coherence time of these temporal windows is limited by the shortest pulse duration in our system ($\tau_{p} = \sim$150 fs\autocite{Gennaro2018_barDiscs_PulseChar}). 
As $2d \gg c\tau_{p}$ and the repetition rate of the laser, 80 MHz, is mismatched with the interferometer arm lengths, $\sim$20 cm, only light arriving within the same temporal window will interfere. 
Therefore, when calculating the intensity at the detector, $D$, fields arriving within the same temporal window are grouped together
\begin{equation} \label{SI_intensity@detector}
 D = |E_{f1-}|^{2} + |E_{f1+}+E_{f2-}|^{2} + |E_{f2+}+E_{f1+}r_{f}e^{2ik_{f}d}|^{2}.
\end{equation}
An additional term has been included due to the back reflection of $E_{f1+}$ off the metasurface leading to a temporal overlap with $E_{f2+}$ at $4k_{f}d$.  
Linear reflections of the pump, seed and FWM will create successive temporal overlap at $mk_{f}d$, where $m$ is an even integer. However, the nonlinear scaling of FWM emission with pump and probe amplitude allows the model to be approximated to the first FWM reflection.
Substituting in the amplitudes in Eqs.~(\ref{SI_f1+_simple}-\ref{SI_f2-_simple}) and normalising the detected signal relative to the forward FWM emission signal, $N=D/\sigma_+^2$, gives us
\begin{equation} \label{SI_Normintensity@detector}
N=R_f T_f+(\eta^2+T_f)(1+\kappa T)+2(\eta+\sqrt{T_f R_f})\sqrt{\kappa TT_f}\cos(\gamma+\psi),
\end{equation}
Where $T = |t|^2$ and $R=|r|^2$, $\kappa=T_p^2 T_s$ is the ratio of total FWM (signal) from the two passes, and $\eta=\sigma_-/\sigma_+$ is the ratio of emission in the backwards and forwards directions. 
Calculating visibility as $V = \frac{N_{max} - N_{min}}{N_{max} - N_{min}} $ gives
\begin{equation} \label{SI_visibility}
 V = \frac{2\sqrt{\kappa T T_f}(\eta+\sqrt{T_fR_f})}{R_fT_f+(\eta^2+T_f)(1+\kappa T)}.
\end{equation}

\section{Mach-Zehnder configuration}

\begin{figure}[ht!]
    \centering
    \includegraphics[width=0.875\linewidth]{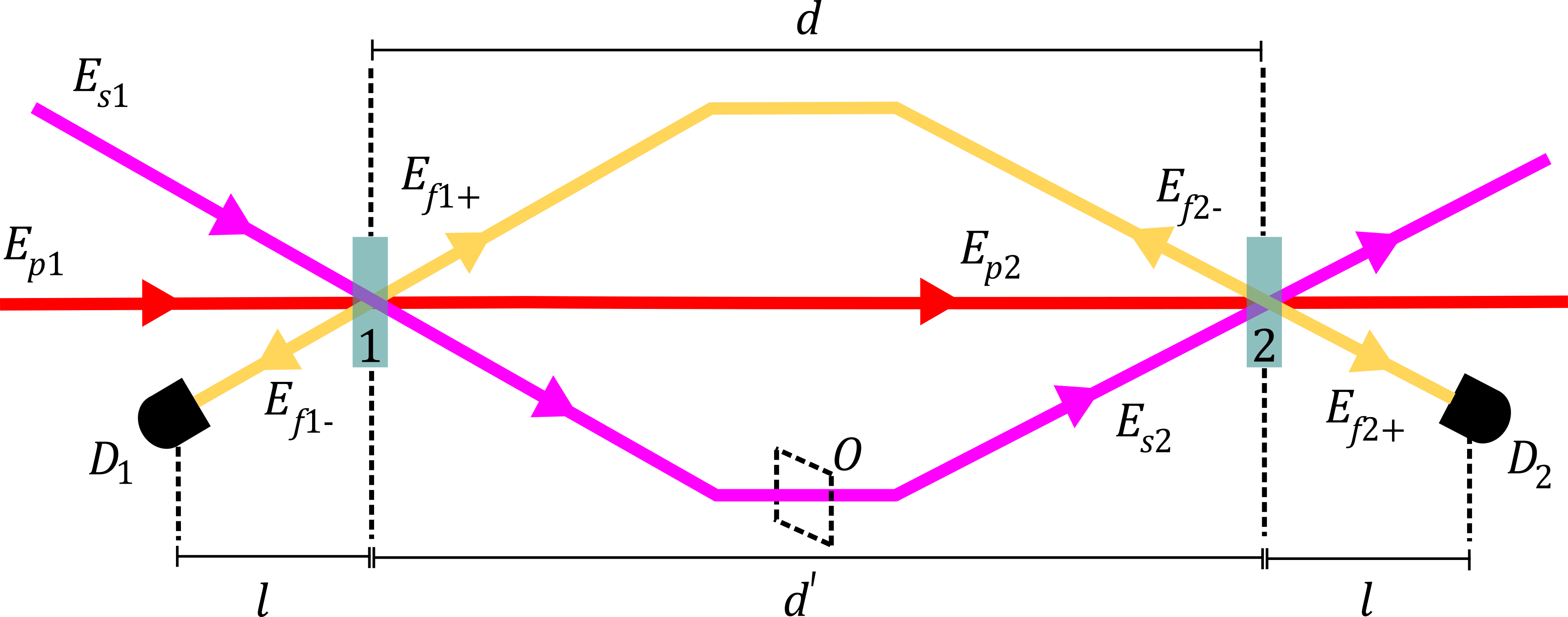}
    \caption{\justifying {\bf A Mach-Zehnder interferometer with two nonlinear metasurface sources}. The colours of the paths mark the pump (red), seed (idler, magenta) and FWM (signal, yellow). The fields $E_{f1\pm,f2\pm}$ represent the bidirectional emission at the first and second source, respectively. Detector $D_1$ measures the backwards emitted FWM fields and detector $D_2$ measures the forward emitted fields. Inside the interferometer, the FWM and pump travel path $d$ and the seed travels path $d^{\prime}$.   }
    \label{fig:resp_Mach-Zehnder}
\end{figure}

In the Michelson style interferometer discussed above and in the main text, one of the four nonlinear fields only contributes to the background, as it is isolated in time from all other fields. However, in a Mach-Zehnder configuration, Fig.~\ref{fig:resp_Mach-Zehnder}, fields are directionally filtered to improve visibility.

In the proposed setup, a pump and seed pass through two identical metasurface or thin-film sources, generating bidirectional emission at each. 
$E_{p1}$ and $E_{s1}$ are the pump and seed fields at the first source, which generate FWM mixing fields $E_{f1+}$ in the forward direction and $E_{f1-}$ in the backward direction. 
$E_{p2}$ and $E_{s2}$ are the pump and seed fields at the second source that generate FWM fields $E_{f2+}$ in the forward direction and $E_{f2-}$ in the backward direction. 
Detector 1, $D_1$, measures the intensity of the backward-generated fields and detector 2, $D_2$, measures the intensity of the forward-generated fields.

The incident fields at the first source are defined as in Eqs.~(\ref{SI_pump1}) and (\ref{SI_seed1}) in the derivation above. 
From the paths labelled in Fig.~\ref{fig:resp_Mach-Zehnder}, the pump and seed fields at the second source are
\begin{equation} \label{MZ_SI_Ep2}
   E_{p2} = t_p E_{p1}e^{i k_p d} \textrm{ and}
\end{equation}
\begin{equation} \label{MZ_SI_Es2}
   E_{s2} = t_s t e^{i\gamma} E_{s1}e^{i k_s d^{\prime}} \textrm{.}
\end{equation}

Here, $t_{p,s}$ are the transmission coefficients of the pump and seed beam amplitudes through the first source, and $t$ and $\gamma$ are the transmission coefficient and phase accrued by the seed at the object plane $O$. 
As we have assumed the two sources are identical, the nonlinear fields generated at each source have the form identified in Eq.~(\ref{SI_FWMstrength}).
As before, we assume the pump and seed are undepleted by the FWM process. From here, we assess the FWM fields at the respective detectors.
\begin{equation} \label{MZ_SI_FWM_Ef1-}
    E_{f1-} = \alpha_- E^2_{p1} E^*_{s1} e^{ik_f l}
\end{equation}
\begin{equation} \label{MZ_SI_FWM_Ef1+}
    E_{f1+} = \alpha_+ t_f E^2_{p1} E^*_{s1} e^{ik_f (d+l)}
\end{equation}
\begin{equation} \label{MZ_SI_FWM_Ef2-}
    E_{f2-} = \alpha_- t_f E^2_{p2} E^*_{s2} e^{ik_f (d+l)} = \alpha_- t_f t_s t^2_p t e^{i\gamma} E^2_{p1} E^*_{s1} e^{i 2d k_p - id^{\prime} k_s} e^{ik_f (d + l)}
\end{equation}
\begin{equation} \label{MZ_SI_FWM_Ef2+}
    E_{f2+} = \alpha_+  E^2_{p2} E^*_{s2} e^{ik_f l} = \alpha_+ t_s t^2_p t e^{i\gamma} E^2_{p1} E^*_{s1} e^{i 2d k_p - id^{\prime} k_s} e^{i k_f l}
\end{equation}

The FWM and pump beams take a path of length $d$ and the seed light takes a path of length $d^{\prime}$, if directed through the interferometer. The conservation of energy and momentum in the FWM process is stated in Eq.~(\ref{SI_E&K_conserve}) and the path difference within the interferometer is defined in Eq.~(\ref{SI_pathDiff}). This path difference produces the phase shift
\begin{equation} \label{MZ_SI_phaseShift}
 k_s\delta = \psi \textrm{.}
\end{equation}
We also collect like terms into one parameter for each direction, that represents the average nonlinear amplitude in Eq.~(\ref{SI_NIprocess_phase}).
These simplifications are substituted into the FWM fields in Eqs.~(\ref{MZ_SI_FWM_Ef1-}-\ref{MZ_SI_FWM_Ef2+}).
\begin{equation} \label{MZ_SI_FWM_Ef1-_simp}
    E_{f1-} = \sigma_-
\end{equation}
\begin{equation} \label{MZ_SI_FWM_Ef1+_simp}
    E_{f1+} = \sigma_+ e^{ik_fd}
\end{equation}
\begin{equation} \label{MZ_SI_FWM_Ef2-_simp}
    E_{f2-} = \sigma_- t_f t_s t^2_p t e^{i\gamma} e^{i \psi} e^{i 2k_f d } 
    \end{equation}
\begin{equation} \label{MZ_SI_FWM_Ef2+_simp}
    E_{f2+} = \sigma_+ t_s t^2_p t e^{i\gamma} e^{i \psi} e^{i k_f d}
\end{equation}

As before, we have separate temporal/phase windows defined by the paths taken by the various fields. 
The two forward generated fields overlap at $k_fd$, while the backward generated fields are separated from one another by $2k_f d$.
We make the assumption that $d \ll cR^{-1}$, where $R$ is the repetition rate of our laser, such that subsequent pulses do not overlap.
However, the $E_{f1+}$ field reflects off source 2 towards detector 1 and is overlapped with the $E_{f2-}$ field. This means at $D_1$ we have the intensity

\begin{equation} \label{MZ_SI_D1_I}
    D_{1} = |E_{f1-}|^2+|E_{f2-}+ r_f e^{ik_f d} E_{f1+}|^2.
\end{equation}

Substituting Eqs.~(\ref{MZ_SI_FWM_Ef1-_simp}-\ref{MZ_SI_FWM_Ef2-_simp}) into Eq.~(\ref{MZ_SI_D1_I}) leads to 

\begin{equation} \label{MZ_SI_D1_I_sub}
    D_{1} = |\sigma_-|^2+|\sigma_- t_f t_s t^2_p t e^{i\gamma}e^{i\psi}e^{i2k_fd}+t_fr_f\sigma_+e^{i2k_fd}|^2.
\end{equation}
Expanding Eq.~(\ref{MZ_SI_D1_I_sub}) and normalising to $\sigma^2_+$ results in
\begin{equation} \label{MZ_SI_D1_I_norm}
    N_{1} = \frac{D_1}{\sigma_+^2} = \eta^2(1+\kappa T_f T)+T_f R_f+ 2\eta T_f\sqrt{\kappa R_f T} \mathrm{cos}(\psi+\gamma).
\end{equation}
Here, $T = |t|^2$ and $R = |r|^2$, $\kappa=T_p^2 T_s$ is the ratio of the total FWM (signal) from the two passes, and $\eta=\sigma_-/\sigma_+$ is the ratio of emission in the backward and forward directions.
The above equation has oscillating and non-oscillating terms, and visibility can be calculated as $V=\frac{N^{max}_1-N^{min}_1}{N^{max}_1+N^{min}_1}$.
\begin{equation} \label{MZ_SI_D1_vis}
    V_{1} = \frac{2\eta T_f\sqrt{\kappa R_f T}}{\eta^2(1+\kappa T_f T)+T_f R_f}
\end{equation}

At $D_2$ we have interference between the forward emitted fields.
\begin{equation} \label{MZ_SI_D2_fields}
    D_{2} = |E_{f1+}+E_{f2+}|^2
\end{equation}

Substituting in the equations for the fields in Eqs.~(\ref{MZ_SI_FWM_Ef1+_simp}) and (\ref{MZ_SI_FWM_Ef2+_simp}) leads to
\begin{equation} \label{MZ_SI_D2_I}
    D_{2} = |\sigma_+ e^{ik_fd} + \sigma_+ t_s t^2_p t e^{i\gamma} e^{i \psi} e^{i k_f d}|^2.
\end{equation}

We expand out the above equation and normalise to the forward emitted nonlinear amplitude.
\begin{equation} \label{MZ_SI_D2_I_norm}
    N_{2} = \frac{D_{2}}{\sigma_+} =T_f +\kappa T+ 2 \sqrt{\kappa T T_f}\mathrm{cos}(\gamma +\psi)
\end{equation}
From this equation, the visibility can be calculated from $V_2=\frac{N^{max}_2-N^{min}_2}{N^{max}_2+N^{min}_2}$.

\begin{equation} \label{MZ_SI_D2_vis}
    V_{2} = \frac{2\sqrt{\kappa T T_f}}{T_f+\kappa T}
\end{equation}

We have arrived at a scenario where $D_1$ measures interference with a visibility limit by the non-interfering field, similar to that of the Michelson configuration in the main text, but $D_2$ measures interference theoretically capable of 100\% visibility, similar to conventional nonlinear interferometers. 
Because the $E_{s2}$ field travels through the object plane to generate $E_{f2+}$, the interference in the backward direction contains transmission information about the object at $O$. 
It is worth noting that the interfering fields that reach $D_1$ will also reflect off source 1 and reach $D_2$, but this should not impact the maximum achievable visibility if all fields are coherent.

\section{Tilted thin-film metasurface source}

\begin{figure}[ht!]
    \centering
    \includegraphics[width=0.65\linewidth]{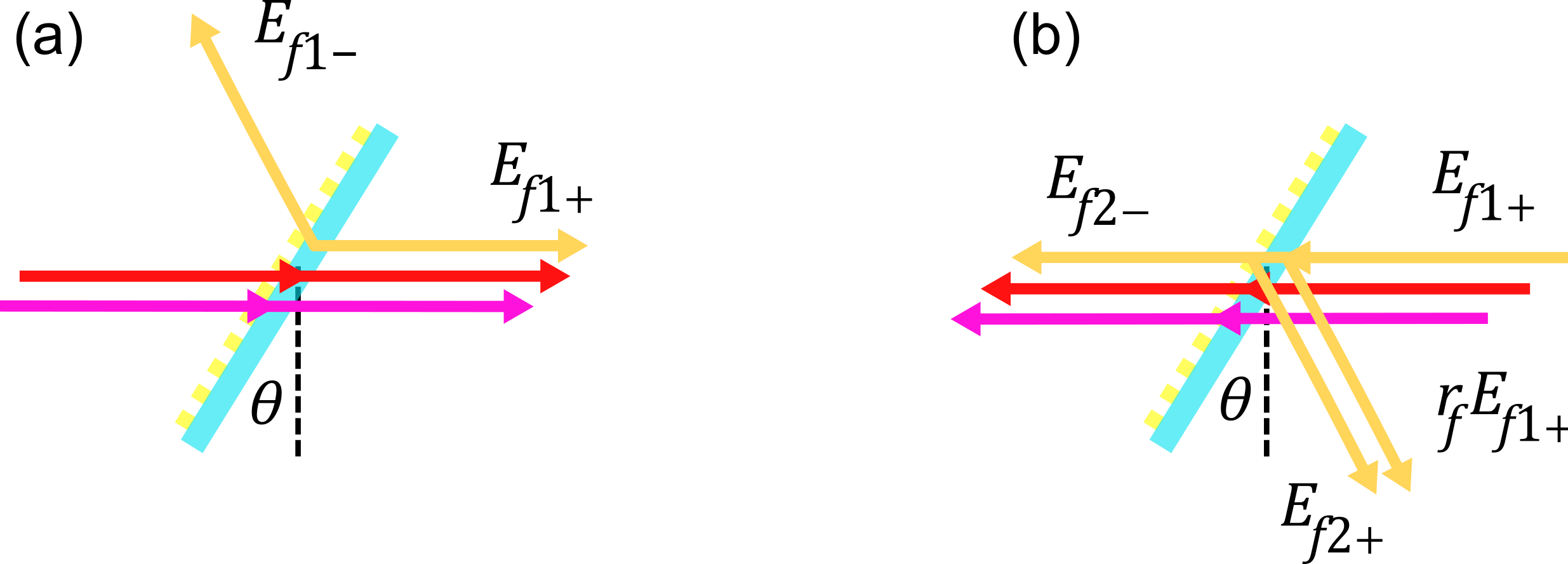}
    \caption{\justifying {\bf Angular misalignment of FWM (signal) fields when the metasurface source is tilted inside a Michelson interferometer}. (a) FWM (signal) fields at the first pass, where the $E_{f1-}$ field is reflected away from the optical axis of the pump and seed. (b) FWM (signal) fields at the second pass, where the secondary interference pattern is reflected away from the optical axis.
    }
    \label{fig_SI_tilted}
\end{figure}

An alternative route to separating the nonlinear fields involves tilting the metasurface source(s). 
This would be effective in a Michelson or Mach-Zehnder configuration; however, it is described here with reference to the former. 
The diagrams in Fig.~\ref{fig_SI_tilted} show the pump, seed (idler) and FWM (signal) fields at the two passes through a tilted metasurface.   

The metasurface is tilted at an arbitrary angle $\theta$ from normal incidence with the optical axis, shared by the collinear pump and seed (idler) beams. 
At the first pass, the FWM (signal) field generated in the positive direction ($E_{f1+}$) is emitted along the optical axis. 
However, the negative FWM (signal) field ($E_{f1-}$) is emitted  $-\theta$ from the sample normal ($2\theta$ from the pump-seed optical axis). 
The emission angle is due to the in-plane momentum conservation of the nonlinear process at the metasurface. 
A similar scenario is expected at the second pass, where $E_{f2-}$ is emitted along the optical axis, but the $E_{f2+}$ field is emitted at $-\theta$. 
The first pass field is reflected from the metasurface ($r_{f} E_{f1+}$) at the same angle. 
Therefore, the tilted metasurface introduces three paths, one for each of the three temporal/phase windows identified in Sec.~\ref{SI_SEC_VisDerivation}. 
As each set of fields is now separated angularly as well as by their phase, we define the nonlinear amplitude and FWM (signal) fields without path information.
\begin{equation} \label{SI_tilt_NL_Amp}
    \alpha_{\pm}E^2_p E^*_s e^{i(2\phi_p(\omega)-\phi_s(\omega))}= \widetilde{\sigma}_{\pm}e^{i\phi_f(\omega)}=\sigma_\pm
\end{equation}
\begin{equation} \label{SI_tilt_Ef1-}
    E_{f1-}=\sigma_-
\end{equation}
\begin{equation} \label{SI_tilt_Ef1+}
    E_{f1=}=\sigma_+
\end{equation}
\begin{equation} \label{SI_tilt_Ef2-}
    E_{f2-}=t t_s t_p^2 \sigma_- e^{i\gamma} e^{i\psi}
\end{equation}
\begin{equation} \label{SI_tilt_Ef2+}
    E_{f2+}=t t_s t_p^2 \sigma_+ e^{i\gamma} e^{i\psi}
\end{equation}

By placing detectors outside the interferometer to terminate each path, the intensity of each set of fields can be measured separately.

\begin{equation} \label{SI_tilt_D1_I}
    D_1 = |E_{f1-}|^2 = \sigma_-^2
\end{equation}
\begin{equation} \label{SI_tilt_D2_I}
    D_2 = |t_f E_{f1+}+E_{f2-}|^2 = |\sigma_+ t_f + t t_s t_p^2 \sigma_- e^{i\gamma}e^{i\psi}|^2
\end{equation}
\begin{equation} \label{SI_tilt_D2_I_norm}
    N_2=\frac{D_2}{\sigma_+^2} = T_f + T T_s T_p^2 \eta^2 + 2\eta\sqrt{T_f T T_s T_p^2} \mathrm{cos}(\gamma + \psi)
\end{equation}
\begin{equation} \label{SI_tilt_D3_I}
    D_3 = |E_{f2+} + r_f E_{f1+}|^2 = |t t_s t_p^2 \sigma_+ e^{i\gamma}e^{i\psi} + \sigma_+r_f|^2 
\end{equation}
\begin{equation} \label{SI_tilt_D3_I_norm}
    N_3=\frac{D_3}{\sigma_+^2} = T T_s T_P^2 + R_f + 2\sqrt{T_s T_p^2 R_f T}\mathrm{cos}(\gamma + \psi)
\end{equation}
In the above equations, $T=|t|^2$ and $R=|r|^2$, $\kappa=T_s T^2_p$ and $\eta=\sigma_-/\sigma_+$, as before. 
It is evident that $D_1$ measures a single field; therefore, there is no interference. 
Two fields are expected at $D_2$ and $D_3$, and the normalised intensity at both detectors contain DC and oscillating terms.
A theoretical visibility is calculated at each detector using $V=\frac{N_{max}-N_{min}}{N_{max}+N_{min}}$, as applied in earlier derivations.
\begin{equation} \label{SI_tilt_D2_vis}
    V_2 = \frac{2\eta \sqrt{\kappa T_f T}}{T_f + \kappa \eta^2 T}
\end{equation}
\begin{equation} \label{SI_tilt_D3_vis}
    V_3 = \frac{2\sqrt{\kappa R_f T}}{R_f + \kappa T}
\end{equation}
Both $V_2$ and $V_3$ are dependent on $T$ and are theoretically capable of reaching 100\% visibility, limited only by the optical properties of the applied metasurface source.
$V_3$ is not dependent on $\eta$ due to both fields being emitted in the positive direction.
The non-interfering nonlinear signal that does not reach the object plane can be simultaneously monitored at $D_1$.

\section{Predicted visibility}
\begin{figure}[ht!]
    \centering
    \includegraphics[width=0.99\linewidth]{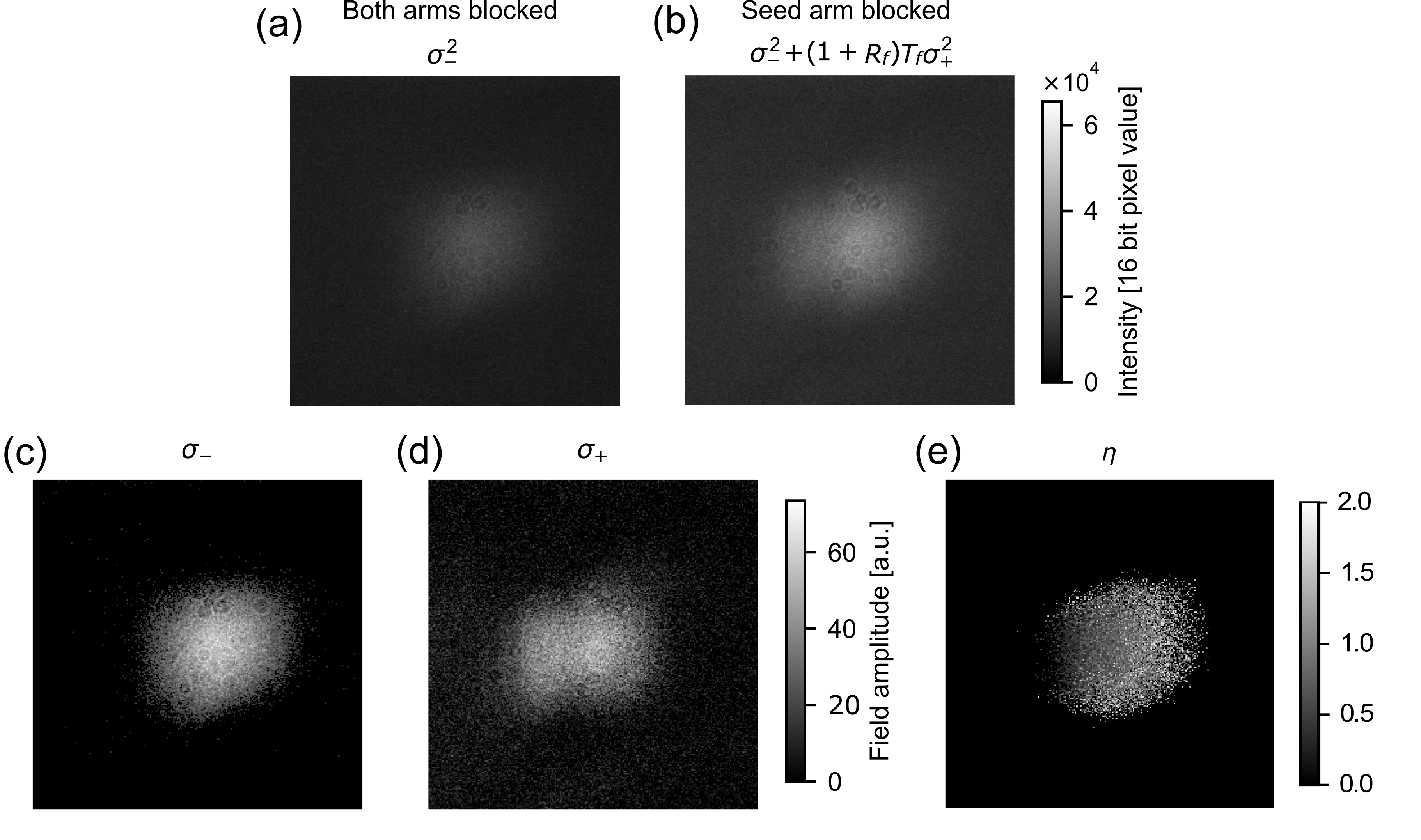} \caption{\justifying{ \bf Extracted amplitudes of bidirectional FWM emission and $\eta$.} 
    (a) Camera response to only first pass reflected emission, $E_{f1-}$. 
    (b) Camera response to the reflected, $E_{f1-}$, and transmitted, $E_{f1+}$, first pass emission. 
    Amplitude of FWM emission at the metasurface in the (c) $-$ and (d) $+$ direction. 
    (e) $\eta$ calculated from (c) and (d). 
    }
    \label{fig:S2}
\end{figure}

\begin{figure}[ht!]
    \centering
    \includegraphics[width=0.75\linewidth]{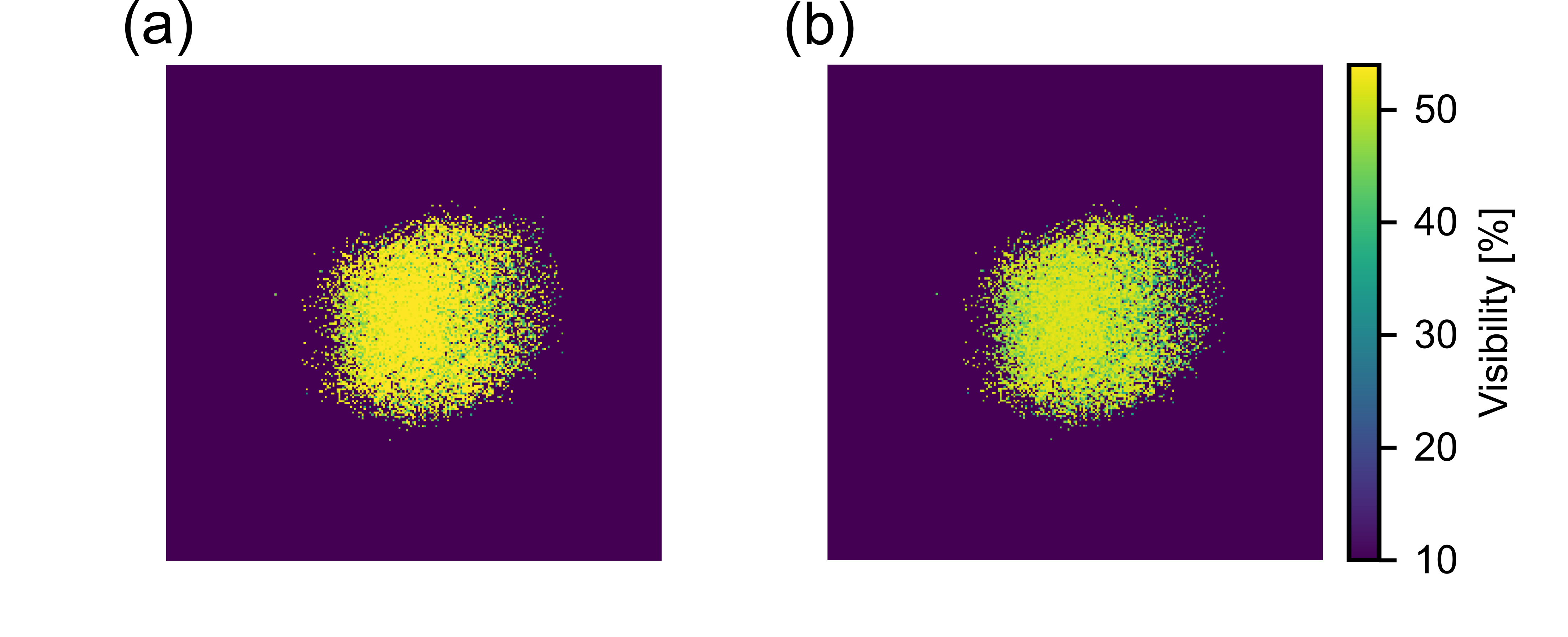} \caption{\justifying{ \bf Expected visibility from extracted $\eta$ and linear optical properties of the metasurface in Table I.} (a) Calculated visibility when T=100\%. (b) Calculated visibility when T=83.3$\pm$1.6\%; the measured transmission of two passes through the 240 \textmu m thick silicon window at 1500 nm.  }
    \label{fig:S3}
\end{figure}

The predicted visibility is calculated from images of the non-interfering fields at the camera when a beam block is placed in the respective arms of the interferometer. 
Fig.~\ref{fig:S2}(a) and (b) show these images labelled with the combination of FWM fields and linear optical properties of the metasurface that define them. 
Far-field images of $\sigma_\pm$ and $\eta$ calculated from Fig.~\ref{fig:S2}(a) and (b) are shown in (c-e). 
The far-field measurement of $\eta$ is then used to predict the visibility across the camera using the parameters of the characterised metasurface to give Fig.~\ref{fig:S3}(a).
The transmission of one pass through the silicon window was measured directly to be 91.3 $\pm$ 0.9\% at 1500 nm, and was subsequently used to calculate Fig.~\ref{fig:S3}(b). 
Here, the increase in optical path length due to the silicon window, which in turn reduces visibility, has not been considered.

\printbibliography